\documentclass[12pt]{article}

\usepackage{graphicx}
\usepackage{amsmath,amssymb}
\usepackage{bm}
\usepackage{graphicx, color}
\usepackage{wrapfig}
\begin{document}
\title{
\begin{flushright}
\ \\*[-80pt] 
\begin{minipage}{0.2\linewidth}
\normalsize
KUNS-2214 \\*[50pt]
\end{minipage}
\end{flushright}
{\Large \bf 
$\Delta(54)$ Flavor Model for Leptons and Sleptons
\\*[20pt]}}

\author{
\centerline{
Hajime~Ishimori$^{1,}$\footnote{E-mail address: 
ishimori@muse.sc.niigata-u.ac.jp}, 
  \quad Tatsuo~Kobayashi$^{2,}$\footnote{E-mail address: 
kobayash@gauge.scphys.kyoto-u.ac.jp},
\quad Hiroshi Okada$^{3,}$
\footnote{E-mail address: HOkada@Bue.edu.eg},
}\\
 \centerline{Yusuke~Shimizu$^{1,}$\footnote{E-mail address: 
shimizu@muse.sc.niigata-u.ac.jp},
\quad and \quad  Morimitsu~Tanimoto$^{4,}$\footnote{E-mail address: 
tanimoto@muse.sc.niigata-u.ac.jp} }
\\*[20pt]
\centerline{
\begin{minipage}{\linewidth}
\begin{center}
$^1${\it \normalsize
Graduate~School~of~Science~and~Technology,~Niigata~University, \\ 
Niigata~950-2181,~Japan } \\
$^2${\it \normalsize 
Department of Physics, Kyoto University, 
Kyoto 606-8502, Japan} \\
$^3${\it \normalsize 
Centre for Theoretical Physics, The British University in Egypt, 
El-Sherouk City, 11837,  Egypt} \\
$^4${\it \normalsize
Department of Physics, Niigata University,~Niigata,  950-2181, Japan } 
\end{center}
\end{minipage}}
\\*[50pt]}
\vskip 2 cm
\date{\small
\centerline{ \bf Abstract}
\begin{minipage}{0.9\linewidth}
\medskip 
We study a $\Delta(54)\times Z_2$ flavor model for leptons and sleptons.
The tri-bimaximal mixing can be reproduced for arbitrary neutrino masses
if certain vacuum alignments of scalar fields are realized. 
The deviation from the tri-bimaximal mixing of leptons is predicted.
The predicted upper bound for $\sin\theta_{13}$  is $0.07$.
The  value of $\sin\theta_{23}$ could be deviated from the maximal mixing
 considerably while $\sin\theta_{12}$  is hardly deviated from
$1/\sqrt3$.
We also study SUSY breaking terms in the slepton sector.
Three families of left-handed and right-handed slepton masses  are almost degenerate.
Our model
leads to smaller values of flavor changing neutral currents 
than the present experimental bounds.
\end{minipage}
}

\begin{titlepage}
\maketitle
\thispagestyle{empty}
\end{titlepage}

\section{Introduction}

Recent experimental data of the neutrino oscillation
indicate the tri-bimaximal form~\cite{HPS} of mixing angles in the 
lepton sector within a good accuracy~\cite{Threeflavors,fogli}.
Thus, it is a promising step to study how to realize the 
tri-bimaximal mixing matrix, in order to understand 
the origin of the lepton flavor.
Many authors have been attempting it by using various scenarios.

Non-Abelian discrete flavor symmetries are particularly well-known as
one of quite verifiable methods  to realize the
tri-bimaximal mixing matrix.
Non-Abelian discrete flavor symmetries 
can provide a natural guidance to constrain
many free parameters in the Yukawa sector. 
Actually, several types of models with various non-Abelian discrete flavor 
symmetries have been proposed, such as 
$S_3$ \cite{S3-FTY}-\cite{S3-Lavoura}, 
 $D_4$ \cite{Grimus}-\cite{Ishimori:2008ns}, $D_6$ \cite{D6},
$Q_4 \cite{Q4},\ Q_6$ \cite{Q6},
$A_4$ \cite{A4-Ma and Rajasekaran}-\cite{A4-Feruglio}, $T^\prime $ 
\cite{T-Feruglio}-\cite{T-Ding}, 
$S_4$ \cite{S4-Ma}-\cite{S4-Bazzocchi2} and 
$\Delta (27)$ \cite{Delta(27)-Gerard}-\cite{Delta(27)-Ma(2008)}.

As another aspect, non-Abelian discrete flavor symmetries could also have
an advantage of supersymmmetry (SUSY) flavor changing neutral 
currents (FCNCs)
\footnote{While they might have a potential disadvantage of FCNC 
through extended Higgs fields, but one could avoid such a problem 
as far as one could stay at the scenarios with $SU(2)_L$ singlet 
extended Higgs fields. }
.
In general, there are a large number of free parameters mainly
related to soft SUSY breaking terms  even in the minimal supersymmetric
standard model.
However, once one could apply 
non-Abelian discrete flavor symmetries, those SUSY breaking parameters 
could be restricted and 
predictable in a way similar to the Yukawa sector. 
(See e.g. 
\cite{Ishimori:2008ns,Kobayashi:2003fh,Ko:2007dz,Ishimori:2008au,vives}.)
Thus, it would also be important to investigate 
non-Abelian discrete flavor symmetries from the viewpoint 
of the SUSY FCNCs.

In addition to the above (rather) bottom-up motivation, 
we also have a top-down motivation.
Certain classes of non-Abelian flavor
symmetries can be derived from superstring theories.
For example, $D_4$ and $\Delta(54)$ flavor symmetries can be obtained
in heterotic orbifold models 
\cite{Ko:2007dz,Kobayashi:2004ya,Kobayashi:2006wq}.
In addition to these flavor symmetries, the $\Delta(27)$ flavor symmetry 
can be derived from magnetized/intersecting D-brane models
\cite{Abe:2009vi}.
Thus, it is quite important to study phenomenological aspects 
of these non-Abelian flavor symmetries.

Here, we focus on the $\Delta(54)$ discrete symmetry 
\cite{Lam,delta54,d54our}. 
Although it includes several interesting 
aspects, few authors have considered up to now. 
The first aspect is that it consists of  
two types of $Z_3$ subgroups and an $S_3$ subgroup.
The $S_3$ group is known as the minimal non-Abelian discrete symmetry, 
and the semi-direct product structure of $\Delta(54)$ 
between $Z_3$ and $S_3$  induces triplet 
irreducible representations.
That suggests that the $\Delta(54)$ symmetry could lead to 
interesting models.

The authors have already presented a $\Delta(54)$ flavor model 
\cite{d54our},
in which the tri-bimaximal mixing
of lepton flavors is reproduced in the vanishing limit  of 
the solar neutrino mass-squared difference. 
Although the previous $\Delta(54)$ model is simple in the sense that the three 
generations of all lepton 
sectors are assigned to be $\Delta(54)$ triplets and it does not need 
additional symmetry such as $Z_n$,
neutrino mass parameters must be  tuned to reproduce 
experimental neutrino data by hand.
 In this paper, we present a new $\Delta(54)$ flavor model,
which is improved to exactly provide the tri-bimaximal matrix 
for arbitrary neutrino masses.
In the present model, 
 the three generations of 
right-handed neutrinos 
are divided into singlet and doublet representations of $\Delta(54)$ 
and the additional $Z_2$ symmetry is imposed for the lepton sector. 

This paper is organized as follows.
In section 2,  our new $\Delta(54)\times Z_2$ lepton flavor model
 is presented.
In section 3  the possible deviation from the tri-bimaximal mixing is discussed,
and 
in section 4  numerical results are presented.
In section 5,  soft SUSY breaking terms of sleptons
 are studied by taking account of FCNC constraints.
Section 6 is devoted to summary and discussion.
In Appendix A,  the analytic derivation of the mixing angles is given, 
and  in Appendix B,  soft SUSY breaking terms in the previous  
$\Delta(54)$ flavor model~\cite{d54our} are  summarized.

\section{$\Delta(54)$ flavor model for leptons}

In this section, we present the lepton flavor model with 
the $\Delta(54)$ flavor symmetry.
We propose a new  model within the framework of supersymmetric theory.
Therefore, we can discuss this flavor symmetry  in the slepton sector
 by constraining parameters of this model  in the lepton sector.

The $\Delta(54)$ group is one of $\Delta(6n^2)$ series
that has been discussed by a few authors \cite{Lam,delta54}. The group
$\Delta(54)$ has irreducible representations $1_1$, $1_2$, $2_1$, $2_2$,
$2_3$, $2_4$, $3_1^{(1)}$, $3_1^{(2)}$, $3_2^{(1)}$,
and $3_2^{(2)}$.
There are four triplets and products of 
$3_1^{(1)}\times 3_1^{(2)}$ and $3_2^{(1)}\times 3_2^{(2)}$ lead to the trivial singlet. 
The relevant  multiplication rules are shown, e.g.  
in Ref.~\cite{Lam,delta54}.

\begin{table}[h]
\begin{center}
\begin{tabular}{|c|cccc||c||cccc|}
\hline
              &$(l_e,l_\mu,l_\tau)$\hspace{-3mm}&\hspace{-3mm} $(e^c,\mu^c ,\tau^c )$  
&\hspace{-3mm}$N_e$&\hspace{-3mm}$(N_\mu ,N_\tau )$       
&$h_{u(d)}$ &\hspace{-3mm}$\chi_1$&\hspace{-3mm}$ (\chi_2,\chi_3)$&\hspace{-3mm}$ (\chi_4,\chi_5 )$
&\hspace{-3mm}$ (\chi_6,\chi_7,\chi_8 )$  
\\ \hline
$\Delta(54)$      
              &$3_1^{(1)}$ &$3_2^{(2)}$ &$1_1$&$2_1$ 
& $1_1$& $1_2$& $2_1$& $2_1$ & $3_1^{(2)}$  \\
$Z_2$      
              &$1$ &$-1$ &$1$&$ 1$ 
& $ 1$& $-1$& $ -1$& $1$ & $1$  \\
\hline
\end{tabular}
\end{center}
\caption{Assignments of $\Delta(54)\times Z_2$ representations}
\end{table}

Here, we present our model of the lepton flavor with the $\Delta(54)$ 
group. 
The triplet representations of the $\Delta(54)$ group
correspond to the three generations of left-handed leptons and 
right-handed charged leptons while right-handed neutrinos are assigned 
to a singlet and a doublet   of $\Delta(54)$. 
The left-handed leptons $(l_e,l_\mu,l_\tau)$,  
the right-handed charged leptons  $(e^c,\mu^c,\tau^c)$
are assigned to be  $3_1^{(1)}$ and $3_2^{(2)}$, respectively. 
For right-handed neutrinos, $N_e^c$ is assigned to be $1_1$ and 
$(N_\mu^c,N_\tau^c)$ are assigned to be $2_1$. 
Charged leptons cannot have mass terms 
unless new scalars $\chi_i$ are introduced  in addition to the usual Higgs
doublets, $h_u$ and $h_d$. 
These new scalars are supposed to be $SU(2)_L$ gauge singlets 
with vanishing $U(1)_Y$ charge.
Gauge singlets $\chi_1$, $(\chi_2, \chi_3)$, $(\chi_4, \chi_5)$ and
$(\chi_6, \chi_7, \chi_8)$ are assigned to be 
$1_2$, $2_1$, $2_1$, and $3_1^{(2)}$, 
respectively. We also introduce $Z_2$ symmetry and the non-trivial charge 
is assigned to $(e^c,\mu^c,\tau^c)$, $\chi_1$ 
and $(\chi_2,\chi_3)$. 
The particle assignments of $\Delta(54)$ and $Z_2$ are summarized in Table 
1.
The usual Higgs doublets $h_u$ and $h_d$ are assigned to  
the trivial singlet $1_1$. 
Here, all fields denote superfields,  and in section 5 
 the superfield and its lowest scalar component are denoted 
 by the same letter as a convention.

In this particle assignment,
we consider the superpotential of leptons at the leading order
in terms of the cut-off scale $\Lambda$, which is taken to be 
the Planck scale.
For charged leptons, the superpotential of 
the Yukawa sector respecting $\Delta(54)$ and $Z_2$ symmetries
is given by 
\begin{eqnarray}
W^{(l)}
&=&
y_1^l 
 ( e^c l_e+ \mu^c l_\mu+ \tau^c l_\tau)\chi _1h _d
\nonumber\\
&&+y_2^l\left [
 (-\omega e^c l_e-\omega^2 \mu^c l_\mu- \tau^c l_\tau)\chi_2
+ ( e^c l_e+\omega^2 \mu^c l_\mu+\omega \tau^c l_\tau )\chi_3\right ]
h_d/\Lambda,
\end{eqnarray}
where $\omega = e^{2\pi i/3}$.
For the  right-handed Majorana neutrinos, we can write the superpotential
as follows:
\begin{eqnarray}
W^{(N)}
&=&
M_1N_e^cN_e^c
+M_2( N_\mu^cN_\tau^c+ N_\tau^cN_\mu^c)
\nonumber\\&&
+y^N( N_\mu^cN_\mu^c\chi_4+ N_\tau^c N_\tau^c\chi_5) .
\end{eqnarray}
The superpotential for the Dirac neutrinos is given in leading order as
\begin{eqnarray}
W^{(D)}
&=&
y_1^D
 N_e^c(l_e \chi _6+l_\mu \chi _7+l_\tau \chi _8) h_u  /\Lambda 
\nonumber\\&&
+y_2^D
\left [
 N_\mu^c (\omega l_e \chi _6+\omega ^2l_\mu \chi _7+l_\tau \chi _8)
+N_\tau ^c(l_e \chi _6+\omega ^2l_\mu \chi _7+\omega l_\tau \chi
_8))\right 
]h_u  /\Lambda  .
\end{eqnarray}

We assume that the scalar fields, $h_{u,d}$ and $\chi_i$, develop 
their vacuum expectation values (VEVs) as follows:
\begin{eqnarray}
\begin{split}
&\left<h_{u(d)}\right>=v_{u(d)},
\
\left<\chi_1\right>=u_1,
\
\left<(\chi_2,\chi_3)\right>=(u_2,u_3),
\
\left<(\chi_4,\chi_5)\right>=(u_4,u_5),
\\
&\left<(\chi_6,\chi_7,\chi_8)\right> =(u_6,u_7,u_8) .
\end{split}
\end{eqnarray}
Then, the charged lepton mass matrix is diagonal:
\begin{eqnarray}
M_l
 = 
y_1^lv_d 
\begin{pmatrix}\alpha_1  & 0 & 0 \\ 
           0  & \alpha_1  &  0  \\
                 0  & 0 & \alpha_1   \\
 \end{pmatrix} 
+y_2^l  v_d 
\begin{pmatrix} \omega\alpha_2-\alpha_3 & 0 & 0 \\ 
               0    & \omega^2\alpha_2-\omega^2\alpha_3 &   0 \\
                 0 & 0 &  \alpha_2-\omega\alpha_3  \\
 \end{pmatrix} .
 \end{eqnarray}
The right-handed Majorana mass matrix is given as 
\begin{eqnarray}
M_N
&=&
\begin{pmatrix}M_1  & 0 & 0 \\ 
               0    & y^N\alpha_4\Lambda & M_2   \\
                 0  & M_2 & y^N \alpha_5 \Lambda  \\
 \end{pmatrix} ,
\end{eqnarray}
and the Dirac mass matrix of neutrinos is 
\begin{eqnarray}
M_D
&=&
y_1^Dv_u
\begin{pmatrix}\alpha_6  & \alpha_7 & \alpha_8 \\ 
               0    & 0  &0    \\
                 0  & 0 &0   \\
 \end{pmatrix} 
 +y_2^Dv_u
\begin{pmatrix}0  & 0 & 0 \\ 
               \omega\alpha_6   & \omega^2\alpha_7 &\alpha_8   \\
                \alpha_6  & \omega^2\alpha_7 &\omega\alpha_8   \\
 \end{pmatrix} ,
\end{eqnarray}
where we denote $\alpha_i=u_i/\Lambda \ (i=1, \cdots ,8)$.
By using the seesaw mechanism $M_\nu = M_D^T M_N^{-1} M_D$, the neutrino
mass matrix can be derived.

At first, we analyze the charged lepton sector. 
Masses are expressed by
\begin{eqnarray}
\left(  \begin{array}{cc}
m_e  \\ 
m_\mu  \\ 
m_\tau  \\ 
\end{array} \right)
=
 v_d
\left(  \begin{array}{ccc}
1&\omega &-1  \\ 
1&\omega^2 &-\omega^2  \\ 
1&1 &-\omega  \\ 
\end{array} \right)
\left(  \begin{array}{cc}
y^l_1\alpha_1  \\ 
y^l_2\alpha _2  \\ 
y^l_2\alpha _3  \\ 
\end{array} \right) .
\end{eqnarray}
In order to estimate  magnitudes of $\alpha_1$,  $\alpha_2$ and  
$\alpha_3$,
we  rewrite them as
\begin{eqnarray}
\left(  \begin{array}{cc}
y^l_1\alpha_1  \\ 
y^l_2\alpha _2  \\ 
y^l_2\alpha _3  \\ 
\end{array} \right)
=
\frac{1 }{3v_d}
\left(  \begin{array}{ccc}
1&1 & 1  \\ 
-\omega-1&\omega & 1   \\ 
-1&-\omega &\omega+1  \\ 
\end{array} \right)
\left(  \begin{array}{cc}
m_e \\ 
m_\mu  \\ 
m_\tau  \\ 
\end{array} \right) ,
\end{eqnarray}
which gives the relation of $|y^l_2\alpha_2| = |y^l_2\alpha_3|$.
Inserting the experimental values of the charged lepton masses
with $v_d\simeq 55$GeV (i.e. $\tan\beta=3$),
we obtain numerical results
\begin{equation}
\begin{pmatrix}
y_1^l \alpha _1 \\
y_2^l \alpha _2 \\
y_2^l \alpha _3 
\end{pmatrix} = \begin{pmatrix}
                         1.14\times 10^{-2} \\                                  
                         1.05\times 10^{-2} e^{0.016 i\pi } \\      

                         1.05\times 10^{-2} e^{0.32 i\pi }   \         

\end{pmatrix}.
\label{alpha123}
\end{equation}
Thus, it is found that  $\alpha_i(i=1,2,3)$ are of ${\cal O}(10^{-2})$
when Yukawa couplings are of ${\cal O}(1)$.

In our model,
the lepton flavor mixing is originated from the structure of the neutrino mass 
matrix. 
To realize the tri-bimaximal mixing, we take 
\begin{eqnarray}
\alpha_5=\omega\alpha_4\ , \qquad \qquad \alpha_6=\alpha_7=\alpha_8 .
\label{alin}
\end{eqnarray}
Now we have
\begin{eqnarray}
\begin{split}
M_\nu
=&
\begin{pmatrix}1  & 1 & 1 \\ 
               1   & \omega  &\omega^2    \\
                 1  & \omega^2 &\omega   \\
 \end{pmatrix} 
 \begin{pmatrix}y_1^D   & 0 & 0 \\ 
                   0    & y_2^D\omega   &0    \\
                   0  & 0  & y_2^D    \\
 \end{pmatrix}
\begin{pmatrix}\frac{1}{M_1}  & 0 & 0 \\ 
    0 & \frac{y^N\omega \alpha_4\Lambda}{(y^N\alpha_4\Lambda  
)^2\omega-M_2^2} 
  &\frac{-M_2 }{(y^N\alpha_4\Lambda )^2\omega-M_2^2}    \\
                   0  & \frac{-M_2}{(y^N\alpha_4\Lambda
                     )^2\omega-M_2^2}   
& \frac{y^N\alpha_4\Lambda }{(y^N\alpha_4\Lambda )^2\omega-M_2^2}     \\
 \end{pmatrix}
\\&\times
 \begin{pmatrix}y_1^D   & 0 & 0 \\ 
                   0    & y_2^D\omega   &0    \\
                   0  & 0  & y_2^D    \\
 \end{pmatrix} 
\begin{pmatrix}1  & 1 & 1 \\ 
               1   & \omega  &\omega^2    \\
                 1  & \omega^2 &\omega   \\
 \end{pmatrix}\alpha _6^2v_u^2 .
\end{split}
\end{eqnarray}
It can be rewritten as
\begin{eqnarray}
\begin{split}
M_\nu
=3c
\begin{pmatrix}
1 & 0 & 0 \\
0 & 1 & 0 \\
0 & 0 & 1
\end{pmatrix} +(a-b-c)
\begin{pmatrix}
1 & 1 & 1 \\
1 & 1 & 1 \\
1 & 1 & 1
\end{pmatrix} + 3b
\begin{pmatrix}
1 & 0 & 0 \\
0 & 0 & 1 \\
0 & 1 & 0
\end{pmatrix},
\end{split}
\end{eqnarray}
where 
\begin{equation}
\label{abc}
a=\frac{(y_1^D)^2}{M_1}\alpha _6^2v_u^2,\quad 
b=\frac{y^N(y_2^D)^2\alpha _4\Lambda }{(y^N\alpha_4\Lambda
  )^2\omega-M_2^2}
\alpha _6^2v_u^2, \quad 
c=\frac{-(y_2^D)^2\omega M2}{(y^N\alpha_4\Lambda )^2\omega-M_2^2}
\alpha _6^2v_u^2.
\end{equation}
As well known, the neutrino mass matrix with the tri-bimaximal mixing 
 is expressed in terms of neutrino mass eigenvalues $m_1$, $m_2$ and $m_3 $ by
\begin{equation}
M_\nu = \frac{m_1+m_3}{2}\begin{pmatrix}
                                         1 & 0 & 0 \\ 
                                         0 & 1 & 0 \\
                                         0 & 0 & 1
           \end{pmatrix}+\frac{m_2-m_1}{3}\begin{pmatrix}
                           1 & 1 & 1 \\
                           1 & 1 & 1 \\
                           1 & 1 & 1
       \end{pmatrix}+\frac{m_1-m_3}{2}\begin{pmatrix}
                           1 & 0 & 0 \\
                           0 & 0 & 1 \\
                           0 & 1 & 0
                   \end{pmatrix}.
\end{equation}
Therefore, our neutrino mass matrix $M_\nu $ gives
 the  tri-bimaximal mixing matrix $U_\text{tri}$ and 
  mass eigenvalues  as follows:
\begin{equation}
U_\text{tri} = \begin{pmatrix}
               \frac{2}{\sqrt{6}} &  \frac{1}{\sqrt{3}} & 0 \\
     -\frac{1}{\sqrt{6}} & \frac{1}{\sqrt{3}} &  -\frac{1}{\sqrt{2}} \\
      -\frac{1}{\sqrt{6}} &  \frac{1}{\sqrt{3}} &   \frac{1}{\sqrt{2}}
         \end{pmatrix},
\qquad m_1=3(b+c),\quad m_2=3a,\quad m_3=3(c-b)\ .
\end{equation}
To compare with experimental values, we reparameterize 
$a=|a|$, $b=|b|e^{i\phi}$, $c=|c|e^{i\theta}$,
then neutrino masses become
\begin{eqnarray}
\begin{split}
|m_1|
=&3\sqrt{|b|^2+|c|^2+2|b||c|\cos(\phi-\theta)},
\\
|m_2|
=&3|a|,
\\
|m_3|
=&3\sqrt{|b|^2+|c|^2-2|b||c|\cos(\phi-\theta)}.
\end{split}
\end{eqnarray}
Mass-squared differences are given as
\begin{eqnarray}
\begin{split}
|m_3|^2-|m_1|^2
&=-36|b||c|\cos (\phi -\theta ),
\\
|m_2|^2-|m_1|^2
&=9(|a|^2
-|b|^2-|c|^2-2|b||c|\cos (\phi -\theta )).
\end{split}
\end{eqnarray}

Considering normal-hierarchical neutrino masses,
we take $|b|\simeq |c|$, $\phi-\theta\simeq\pi$, 
then, we get
\begin{eqnarray}
m_1\simeq 0,\quad
m_2\simeq 3|a|,\quad
m_3\simeq  -6|b|e^{i\phi}.
\end{eqnarray}
Parameters $a$ and $b$ are estimated as
$|a|\simeq {\sqrt{\Delta m_\text{sol}^2}}/{3}$, 
$|b|\simeq  {\sqrt{\Delta m_\text{atm}^2}}/{6}$. 
which give the following relations by using of Eq. (\ref{abc}):
\begin{eqnarray}
\begin{split}
&\frac{(y_1^D)^2}{M_1}
v_u^2\alpha_6^2
\simeq\frac{\sqrt{\Delta m_\text{sol}^2}}{3},
\\
&\frac{y^N(y_2^D)^2\alpha_4\Lambda}
{(y^N\alpha_4\Lambda)^2\omega-M_2^2}
v_u^2\alpha_6^2
\simeq
\frac{(y_2^D)^2\omega M_2}
{(y^N\alpha_4\Lambda)^2\omega-M_2^2}
v_u^2\alpha_6^2
\simeq\frac{\sqrt{\Delta m_\text{atm}^2}}{6}
e^{i\phi}.
\label{condition}
\end{split}
\end{eqnarray}
Assuming $\alpha_4$ and $\alpha_6$ to be real, 
the Majorana phase of $m_3$ can be evaluated as
\begin{eqnarray}
e^{i\phi}
\simeq \frac{y^N(y_2^D)^2}{|y^N||y_2^D|^2}
\frac{\sqrt{|y^N|^4\alpha_4^4\Lambda^4
+|M_2|^4-(\omega(y^N)^2{M_2^*}^2+\omega^2({y^N}^*)^2{M_2}^2)
 \alpha_4^2\Lambda^2}}
{\omega(y^N)^2\alpha_4^2\Lambda^2-M_2^2}.
\end{eqnarray}
Because the second equation of (\ref{condition}) implies
$y^N\alpha_4\Lambda\simeq \omega M_2$, 
we set
\begin{eqnarray}
y^N\alpha_4\Lambda=(1+\epsilon) \omega M_2 \ ,
\label{tuning}
\end{eqnarray}
where $\epsilon$ is tiny. 
By using the last equation in Eq. (\ref{condition}), 
the atmospheric neutrino mass scale becomes
\begin{eqnarray}
\begin{split} 
\sqrt{\Delta m_\text{atm}^2}
\simeq \frac{3|y^N||y_2^D|^2\alpha _4\Lambda v_u^2\alpha _6^2}
{\epsilon |M_2|^2} .
\end{split}\label{atm}
\end{eqnarray}
Now we can obtain magnitudes of $\alpha_4$, $\alpha_6$, 
and $M_1$ from experimental values, Yukawa couplings, 
cut-off scale $\Lambda $, Higgs VEVs, right-handed Majorana scale $M_2$, 
and small parameter $\epsilon$. 
Concretely, let us take $v_u=165$GeV, $\Lambda=2.4\times10^{18}$GeV, 
$\Delta m_\text{atm}=2.4\times10^{-21}\text{GeV}^2$, 
$\Delta m_\text{sol}=8\times10^{-23}\text{GeV}^2$. 
Further, putting $|y_1^D|=|y_2^D|=|y^N|=1$, 
$\epsilon=10^{-2}$, and $M_2=10^{13}$GeV, 
we obtain typical values:
\begin{eqnarray}
\begin{split}
\alpha_4
&=\left|\frac{(1+\epsilon)\omega M_2}{y^N\Lambda}\right|
\sim 4\times 10^{-6},
\\
\alpha _6&\sim
\sqrt{
\frac{\epsilon |M_2|^2
\sqrt{\Delta m_\text{atm}^2} }
{3|y^N||y_2^D|^2\alpha_4 \Lambda v_u^2}}\sim 8\times 10^{-3},
\\
M_1&\sim
\frac{3|y_1^D|^2v_u^2\alpha_6^2}
{\sqrt{\Delta m_\text{sol}^2}}
\sim 5\times 10^{11} {\rm GeV}.
\label{numval}
\end{split}
\end{eqnarray}
In this way, we can estimate the magnitudes of 
$\alpha_4$ and $\alpha_6$, which are important parameters to calculate FCNC. 
Even if we consider the case of the inverted mass hierarchy, 
we easily find  the almost same result of $\alpha_4$ and $\alpha_6$ 
 as in Eq. (\ref{numval}). 
In the  section 4, we estimate them numerically
 by taking into account experimental data.   

\section{Deviation from tri-bimaximal mixing}

The tri-bimaximal mixing can be exactly obtained 
under the condition of vacuum alignment in Eq.(\ref{alin}). 
The mixing matrix is deviated from the tri-bimaximal matrix 
if the alignment of Eq.(\ref{alin}) is shifted.

First, we discuss the effect of the deviation from
 $\alpha_5=\omega\alpha_4$. 
To estimate this effect, 
we introduce a parameter $\delta$ 
with $\alpha_5=\omega(1+\delta)\alpha_4$. 
The neutrino mass matrix is written as
\begin{equation}
M_\nu
=
\begin{pmatrix}1  & 1 & 1 \\ 
               1   & \omega  &\omega^2    \\
                 1  & \omega^2 &\omega   \\
 \end{pmatrix} 
 \begin{pmatrix}a   & 0 & 0 \\ 
                   0    & (1+\delta) b   &c    \\
                   0  & c & b  \\
 \end{pmatrix}
\begin{pmatrix}1  & 1 & 1 \\ 
               1   & \omega  &\omega^2    \\
                 1  & \omega^2 &\omega   \\
 \end{pmatrix},
\end{equation}
which is rewritten as
\begin{eqnarray}
\begin{split}
M_\nu
=3c
\begin{pmatrix}
1 & 0 & 0 \\
0 & 1 & 0 \\
0 & 0 & 1
\end{pmatrix} +(a-b-c)
\begin{pmatrix}
1 & 1 & 1 \\
1 & 1 & 1 \\
1 & 1 & 1
\end{pmatrix} + 3b
\begin{pmatrix}
1 & 0 & 0 \\
0 & 0 & 1 \\
0 & 1 & 0
\end{pmatrix}
+ b\delta
\begin{pmatrix}1   &  \omega &  \omega^2 \\ 
                    \omega   & \omega^2   &1    \\
                    \omega^2  & 1  &  \omega    \\
 \end{pmatrix},
\end{split}
\label{Matrix1}
\end{eqnarray}
where the last matrix in the right hand side causes the deviation from
the tri-bimaximal mixing.
It can be diagonalized by the following mixing matrix 
\begin{eqnarray}
U=
\begin{pmatrix}\frac{\sqrt2e^{-ip_1}}{\sqrt3}\cos\theta  & \frac{1}{\sqrt3} 
& -\frac{\sqrt2e^{-ip_1}}{\sqrt3}\sin\theta \\ 
              -\frac{e^{-ip_1}\cos\theta+\sqrt3e^{-ip_2}\sin\theta}{\sqrt6}  
  & \frac{1}{\sqrt3}   
 &\frac{e^{-ip_1}\sin\theta-\sqrt3e^{-ip_2}\cos\theta}{\sqrt6}   \\
                
-\frac{e^{-ip_1}\cos\theta-\sqrt3e^{-ip_2}\sin\theta}{\sqrt6} 
 & \frac{1}{\sqrt3} 
& \frac{e^{-ip_1}\sin\theta+\sqrt3e^{-ip_2}\cos\theta}{\sqrt6}      \\
 \end{pmatrix},
\label{MNS1}
\end{eqnarray}
where the phase difference $p_1-p_2$ and additional mixing angle $\theta$ 
can be expressed by $m_1$, $m_3$, $b$, and $\delta$ as shown in appendix A. 
Then the lepton mixing matrix element $U_{e3}$ 
can be estimated from these parameters. 
On the other hand, the element $U_{e2}$ does not shift from $1/\sqrt{3}$.
Numerical results are discussed in the next section.

We also consider the deviation from the alignment 
$\alpha_6 = \alpha_7 = \alpha_8$. 
New small parameters $\delta_1$ and $\delta_2$ are added 
to be $\alpha_7=(1+\delta_1)\alpha_6$ and 
$\alpha_8=(1+\delta_2)\alpha_6$. 
Then, we obtain 
\begin{eqnarray}
M_\nu
=
 \begin{pmatrix}1   & 0 & 0 \\ 
                   0    & 1+\delta_1   &0    \\
                   0  & 0  & 1+\delta_2    \\
 \end{pmatrix}
U_\text{tri}
 \begin{pmatrix}m_1   & 0 & 0 \\ 
                   0    &m_2   &0    \\
                   0  & 0 & m_3  \\
 \end{pmatrix}
U_\text{tri}^\dagger
 \begin{pmatrix}1   & 0 & 0 \\ 
                   0    & 1+\delta_1   &0    \\
                   0  & 0  & 1+\delta_2    \\
 \end{pmatrix}.
\label{Matrix2}
\end{eqnarray}
It can be diagonalized by
\begin{eqnarray}
U\simeq  U_\text{tri}
 \begin{pmatrix}\cos\theta_{12}   & \sin\theta_{12} & 0 \\ 
                   -\sin\theta_{12}    & \cos\theta_{12}   &0    \\
                   0  & 0  & 1    \\
 \end{pmatrix}
 \begin{pmatrix}\cos\theta_{13}   & 0 & \sin\theta_{13} \\ 
                   0    & 1   &0    \\
                   -\sin\theta_{13}  & 0  & \cos\theta_{13}    \\
 \end{pmatrix}
  \begin{pmatrix}1   & 0 & 0 \\ 
                   0    & \cos\theta_{23}   &\sin\theta_{23}    \\
                   0  & -\sin\theta_{23}  & \cos\theta_{23}   \\
 \end{pmatrix},
\end{eqnarray}
where
\begin{eqnarray}
\begin{split}
\theta_{12}
&\simeq -\frac{m_1+m_2}{3\sqrt2(m_2-m_1)}
(\delta_1+\delta_2),
\quad
\theta_{13}
\simeq \frac{m_1+m_3}{2\sqrt3(m_3-m_1)}
(\delta_1-\delta_2),
\\
\theta_{23}
&\simeq -\frac{m_2+m_3}{\sqrt6(m_3-m_2)}
(\delta_1-\delta_2).
\end{split}
\end{eqnarray}
Supposing the  normal hierarchy of  neutrino masses
$m_3\gg m_2,m_1$, we find
\begin{eqnarray}
\begin{split}
 U_{e3}
&\simeq
\frac{\sqrt2 m_2 }{3 m_3 }
(\delta_2-\delta_1),
\\
U_{\mu 2}
&\simeq
-\frac{1}{\sqrt2}+\frac{\sqrt2}{4}
(\delta_2-\delta_1).
\end{split}
\end{eqnarray}
Since $U_{e3}$ is strongly suppressed 
by order $\delta_i$ and the ratio $m_2/m_3$, 
we consider no more the deviation of $\alpha_6=\alpha_7=\alpha_8$ 
in our numerically work.

\section{Numerical results}

In this section, we discuss the magnitude of the deviation from 
the tri-bimaximal matrix numerically. 
By restricting neutrino masses and mixing angles within 
experimental errors,  magnitudes of $\alpha_4$ and 
$\alpha_6$ can be obtained  as discussed in the section 3. 
We consider the case of the normal hierarchy of neutrino masses 
\footnote{We have not presented  the case of the inverted hierarchy
of neutrino masses since numerical results are almost same as
ones  in the case of the normal hierarchy.}
as discussed in
section 2. 

Input data of masses and mixing angles are taken in the  region of 
 3$\sigma$ of the experimental data \cite{Threeflavors,fogli}:
\begin{eqnarray}
&&\Delta m_{\rm atm}^2=(2.07\sim 2.75)\times 10^{-3} 
{\rm eV}^2 \ ,
\quad \Delta m_{\rm sol}^2= (7.05\sim 8.34) \times 10^{-5} {\rm eV}^2  \ , 
\nonumber \\
&& \sin^2 \theta_{\rm atm}=0.36\sim 0.67 \ ,
\quad  \sin^2 \theta_{\rm sol}=0.25 \sim 0.37  \ , \quad
 \sin^2 \theta_{\rm reactor} \leq 0.056\ .
\end{eqnarray}
Yukawa couplings $y_1^D$ and $y_2^D$ are complex. Those absolute values
and phases are   chosen  from $-1$ to $1$ and   $0$ to $2\pi$ at random,
respectively. On the other hand, $y^N$ is given in Eq.(\ref{tuning}).
We search the experimentally allowed region 
by diagonalizing the neutrino mass matrix with 
varying the parameters $\alpha_4$, $\alpha_6$, $M_2$, 
and $\epsilon$ in Eq.(\ref{tuning}), which are taken to be real.

\begin{figure}[h]
\includegraphics[width=7cm,height=5cm]{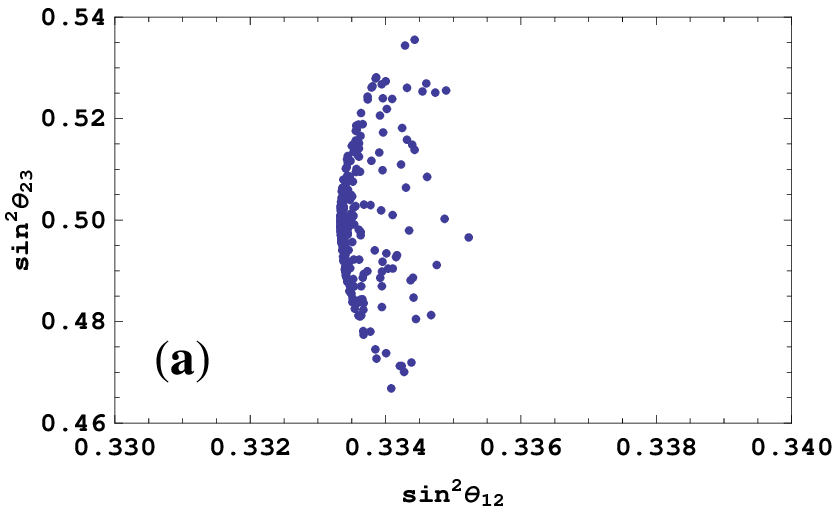}
\quad
\includegraphics[width=7cm,height=5cm]{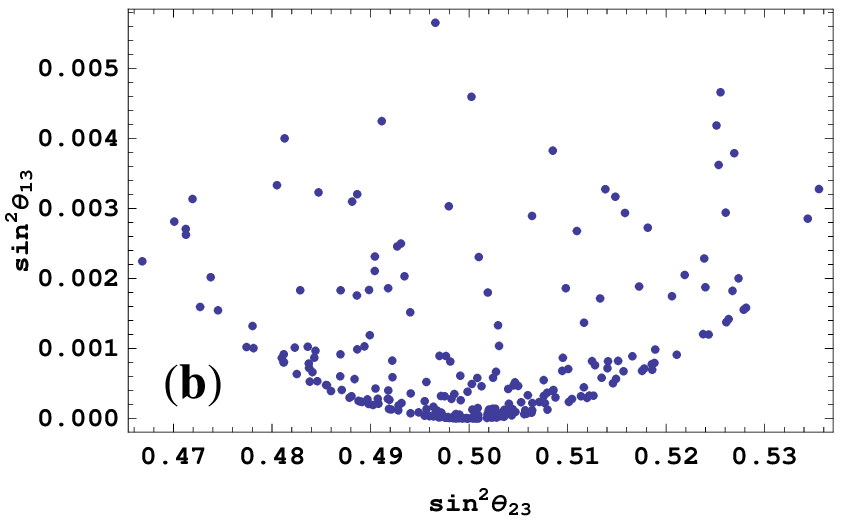}
\end{figure}
\begin{figure}[h]
\includegraphics[width=7cm,height=5cm]{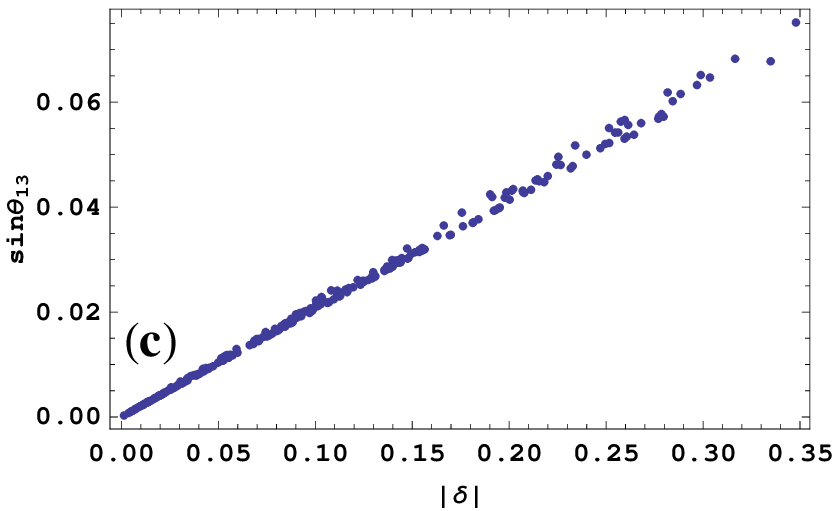}
\quad
\includegraphics[width=7cm,height=5cm]{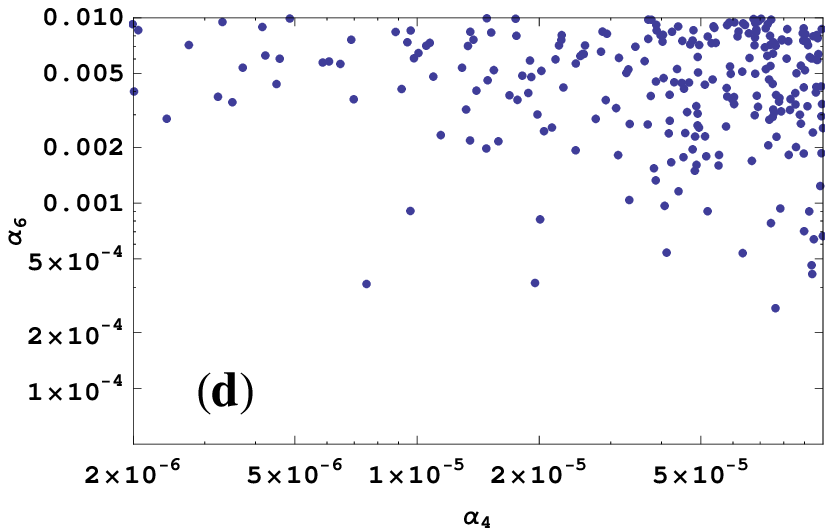}
\caption{Predicted  plots on the plane of
 (a)  $\sin^2 \theta_{23}-\sin^2 \theta_{12}$, 
(b)  $\sin^2 \theta_{23}-\sin^2 \theta_{13}$,
(c)$|\delta|-\sin\theta_{13}$ and (d) $\alpha_4-\alpha_6$ 
 in  the region of   $\alpha_6 \leq 10^{-2}$.}
\end{figure}

As discussed in section 3, we take  $\alpha_5=\omega \alpha_4 (1+\delta)$, 
where $\delta$ is a complex parameter, while we take $\alpha_6=\alpha_7=\alpha_8$. 
By varying $\delta$, the mixing matrix is deviated from 
 the tri-bimaximal matrix.
The neutrino mass matrix is given by 
\begin{equation}
M_\nu
=
\begin{pmatrix}1  & 1 & 1 \\ 
               1   & \omega  &\omega^2    \\
                 1  & \omega^2 &\omega   \\
 \end{pmatrix} 
 \begin{pmatrix}a   & 0 & 0 \\ 
                   0    &b(1+\delta)  &c    \\
                   0  & c & b  \\
 \end{pmatrix}
\begin{pmatrix}1  & 1 & 1 \\ 
               1   & \omega  &\omega^2    \\
                 1  & \omega^2 &\omega   \\
 \end{pmatrix}.
\end{equation}

 We present the numerical result in Figure 1.
In Figure 1(a), we show the allowed region on the plane of 
$\sin^2 \theta_{23}-\sin^2 \theta_{12}$. The mixing parameter 
$\sin \theta_{12}$ is hardly  deviated from the tri-maximal mixing $1/\sqrt{3}$
 as expected from  Eq.(\ref{MNS1}).
On the other hand,  $\sin \theta_{23}$ is 
deviated from the bi-maximal mixing
considerably.  We show allowed values of  $\sin^2 \theta_{13}$
versus  $\sin^2 \theta_{23}$ in Figure 1(b).
The predicted  upper bound of $\sin^2 \theta_{13}$ 
is $5\times 10^{-3}$.   As seen in Figure 1(c),
$\sin\theta_{13}$ is proportional to
the magnitude $|\delta|$, which is bounded by $0.3$ due to the
experimental data of the neutrino mass-squared differences.
In Figure 1(d), we show the allowed region on the $\alpha_4-\alpha_6$ 
plane.
Since $\alpha_6$ larger than  $10^{-2}$ is dangerous 
 for  the  FCNC constraints as discussed in the next section, 
we have searched   the parameter space in $\alpha_6\leq 10^{-2}$.  
Then we find $\alpha_4=10^{-6}\sim 10^{-4}$.
In these numerical calculations, we take 
 $10^{9} {\rm GeV}<M_2<10^{16}{\rm GeV}$ and $\epsilon= 10^{-3}\sim 10^{-1}$.

\section{SUSY breaking terms}

In this section, we study SUSY breaking terms, i.e., 
sfermion masses and scalar trilinear couplings, which 
are predicted in our $\Delta(54)\times Z_2$ model.
We consider the gravity mediation within the 
framework of supergravity theory.
We assume that 
non-vanishing F-terms of gauge and flavor singlet (moduli) fields $Z$ 
and gauge singlet fields $\chi_i$ $(i=1,\cdots,8)$ 
contribute to SUSY breaking.
Their F-components are written as 
\begin{eqnarray}
F^{\Phi_k}= - e^{ \frac{K}{2M_p^2} } K^{\Phi_k \bar{I} } \left(
  \partial_{\bar{I}} \bar{W} + \frac{K_{\bar{I}}} {M_p^2} \bar{W} \right) ,
\label{eq:F-component}
\end{eqnarray}
where $K$ denotes the K\"ahler potential, $K_{\bar{I}J}$ denotes 
second derivatives by fields, 
i.e. $K_{\bar{I}J}={\partial}_{\bar{I}} \partial_J K$
and $K^{\bar{I}J}$ is its inverse. 
Here the fields ${\Phi_k}$ correspond to the moduli fields $Z$ and 
gauge singlet fields $\chi_i$ $(i=1,\cdots,8)$.
The VEVs of $F_{\Phi_k}/\Phi_k$  are estimated as 
$\langle F_{\Phi_k}/ \Phi_k \rangle = {\cal O}(m_{3/2})$, where
$m_{3/2}$ denotes the gravitino mass, which is obtained as 
$m_{3/2}= \langle e^{K/2M_p^2}W/M_p^2 \rangle$.

\subsection{Slepton mass matrices}

First, let us study soft scalar masses.
Within the framework of supergravity theory,
soft scalar mass squared is obtained as \cite{Kaplunovsky:1993rd}
\begin{eqnarray}
m^2_{\bar{I}J} K_{{\bar{I}J}}= m_{3/2}^2K_{{\bar{I}J}} 
+ |F^{\Phi_k}|^2 \partial_{\Phi_k}  
\partial_{  \bar{\Phi_k} }  K_{\bar{I}J}-
|F^{\Phi_k}|^2 \partial_{\bar{\Phi_k}}  K_{\bar{I}L} \partial_{\Phi_k}  
K_{\bar{M} J} K^{L \bar{M}}.
\label{eq:scalar}
\end{eqnarray}

The invariance under the $\Delta(54) \times Z_2$ flavor symmetry 
as well as the gauge invariance requires the following form 
of the K\"ahler potential of $l_I$ and $e_I$ $(I=e,\mu,\tau)$
\begin{equation}
\label{eq:Kahler}
K = Z^{(L)}(Z)\sum_{I=e,\mu,\tau} |l_I|^2 + 
Z^{(R)}(Z)\sum_{I=e,\mu,\tau} |e_I|^2 ,
\end{equation}
at the lowest level, where $Z^{(L)}(Z)$ and $Z^{(R)}(Z)$ are 
arbitrary functions of the singlet fields $Z$.
By use of the formula (\ref{eq:scalar}) with 
the K\"ahler potential (\ref{eq:Kahler}), 
we obtain the following matrix form 
of soft scalar masses squared for left-handed and 
right-handed charged sleptons,
\begin{eqnarray}
(m_{\tilde L}^2)_{IJ} =
\left(
  \begin{array}{ccc}
m_{L}^2   &  0 &  0 \\ 
0   & m_{L}^2  & 0  \\ 
0   &  0  & m_{L}^2   \\ 
\end{array} \right),
\qquad
(m_{\tilde R}^2)_{IJ} 
 = 
\left(
  \begin{array}{ccc}
m_{R}^2   &  0 &  0 \\ 
0   &  m_{R}^2 & 0  \\ 
0   & 0   & m_{R}^2   \\ 
\end{array} \right).
\label{eq:soft-mass-1}
\end{eqnarray}
That is, both matrices are proportional 
to the $(3\times 3)$ identity matrix.
This form would be obvious because 
$(l_e,l_\mu,l_\tau)$ and $(e^c,\mu^c,\tau^c)$ 
are $\Delta(54)$ triplets.
At any rate, it is the prediction of our model that 
three families of left-handed and right-handed masses 
are degenerate.

The above prediction holds exactly before $\Delta(54) \times Z_2$ 
is broken, but its breaking would change the form.
Next, we study effects due to  $\Delta(54) \times Z_2$  breaking 
by $\chi_i$.
That is, we estimate corrections to the K\"ahler potential 
including  $\chi_i$.
The VEVs of $\chi_{4,5}$ are much smaller than the others.
Thus, we concentrate on corrections including $\chi_i$ 
with $i=1,2,3,6,7,8$.

In our model, the left-handed charged leptons 
$(l_e,l_\mu,l_\tau)$ are assigned to be $3^{(1)}_1$ and 
its conjugate representation is $3^{(2)}_1$.
Their multiplication rule is written as 
\begin{eqnarray}
3^{(1)}_1 \times 3^{(2)}_1 = 1_1 + 2_1 + 2_2 + 2_3 +2_4,
\end{eqnarray} 
and that is written more explicitly in terms of elements as 
\begin{eqnarray}
(x_1,x_2,x_3)_{3^{(1)}_1}\times (y_1,y_2,y_3)_{3^{(2)}_1} &  = &  
(x_1y_1+x_2y_2+x_3y_3)_{1_1} \nonumber \\
 &  + & (x_1y_1+\omega^2x_2y_2 + \omega x_3 y_3, 
\omega x_1y_1+\omega^2x_2y_2 + x_3 y_3)_{2_1} \nonumber \\
 &  + & (x_1y_2+\omega^2x_2y_3 + \omega x_3 y_1, 
\omega x_1y_3+\omega^2x_2y_1 + x_3 y_2)_{2_2} \nonumber \\
&  + & (x_1y_3+\omega^2x_2y_1 + \omega x_3 y_2, 
\omega x_1y_2+\omega^2x_2y_3 + x_3 y_1)_{2_3} \nonumber \\
&  + & (x_1y_3+x_2y_1 + x_3 y_2, 
x_1y_2+x_2y_3 + x_3 y_1)_{2_4} .
\end{eqnarray} 
By use of this multiplication rule, we can find that 
linear terms of $\chi_i$  for $i=1,2,3,6,7,8$
do not appear in corrections 
of the K\"ahler potential (\ref{eq:Kahler}).
Although linear terms of $\chi_{4,5}$ can appear in 
diagonal elements of K\"ahler metric, 
those corrections are not important as said above.
Thus, let us estimate  corrections including 
$\chi_i \chi_k$ as well as $\chi_i \chi_k^*$ for $i,k=1,2,3,6,7,8$.
The $\Delta(54) \times Z_2$ flavor symmetric invariance allows 
only the terms such as $\chi_i \chi_k^*$ for $i,k=6,7,8$ to appear 
in off-diagonal entries of the  K\"ahler metric of
$(l_e,l_\mu,l_\tau)$.
For example, the (1,2) entry of the  K\"ahler metric would 
have correction terms like e.g. 
\begin{eqnarray}
\Delta K = \frac{K'(Z)}{\Lambda^2}\chi_i \chi_k^* l_1 l^*_2 + \cdots,
\end{eqnarray}
where $K'(Z)$ is an arbitrary function of $Z$.
On the other hand, the terms such as $\chi_i \chi_k^*$ 
for $i,k=1,2,3$ can appear in the diagonal entries, but 
such corrections only change the overall factor of the form 
K\"ahler potential (\ref{eq:Kahler}).
When we take into account  the corrections from $\chi_i \chi_k^*$ for
$i,k=6,7,8$ to the K\"ahler potential, 
the soft scalar masses squared for left-handed charged sleptons 
have the following corrections, 
\begin{eqnarray}
(m_{\tilde L}^2)_{IJ} = m_{L}^2 
\left(
  \begin{array}{ccc}
1 + {\cal O}(\alpha_6^2)  &  {\cal O}(\alpha_6^2) &  {\cal O}(\alpha_6^2) 
\\ 
{\cal O}(\alpha_6^2)  & 1+ {\cal O}(\alpha_6^2)   & {\cal O}(\alpha_6^2)  
\\ 
{\cal O}(\alpha_6^2)   &  {\cal O}(\alpha_6^2)  & 1 + {\cal O}(\alpha_6^2)   
\\ 
\end{array} \right).
\label{eq:soft-mass-2-L}
\end{eqnarray}
Similarly, when we include the same level of corrections, 
the soft scalar masses squared for right-handed sleptons are 
obtained as 
\begin{eqnarray}
(m_{\tilde R}^2)_{IJ} = m_{R}^2 
\left(
  \begin{array}{ccc}
1 + {\cal O}(\alpha_6^2)  &  {\cal O}(\alpha_6^2) &  {\cal O}(\alpha_6^2) 
\\ 
{\cal O}(\alpha_6^2)  & 1+ {\cal O}(\alpha_6^2)   & {\cal O}(\alpha_6^2)  
\\ 
{\cal O}(\alpha_6^2)   &  {\cal O}(\alpha_6^2)  & 1 + {\cal O}(\alpha_6^2)   
\\ 
\end{array} \right).
\label{eq:soft-mass-2-R}
\end{eqnarray}
These deviations may not be important for 
direct measurement of slepton masses.
However, the off-diagonal entries in the SCKM basis \footnote{
The SCKM basis is the basis, where fermion mass matrix is 
diagonal.} 
are constrained by the FCNC experiments~\cite{FCNCbound}.
Our model predicts
\begin{eqnarray}
(\Delta_{LL})_{12} = \frac{(m_L^2)_{12}^{(SCKM)}}{(m_L^2)_{11}}= {\cal
  O}(\alpha_6^2), 
\qquad 
(\Delta_{RR})_{12} = \frac{(m_R^2)_{12}^{(SCKM)}}{(m_R^2)_{11}}= {\cal
  O}(\alpha_6^2).
\end{eqnarray} 
Recall that the diagonalizing matrices of left-handed and right-handed 
fermions are almost the identity matrix.
The $\mu \rightarrow e \gamma$ experiments constrain 
these values as $(\Delta_{LL,RR})_{12} \leq {\cal
  O}(10^{-3})$~\cite{FCNCbound}, when $m_{L,R} = 100$ GeV.
On the other hand, the parameter space in the previous section 
corresponds to $\alpha_6 \leq 10^{-2}$ and 
leads to $(\Delta_{LL,RR})_{12} \leq {\cal  O}(10^{-4})$.
Thus, our parameter region would be favorable also from 
the viewpoint of the FCNC constraints.

\subsection{A-terms}

Here, let us study scalar trilinear couplings, i.e. 
the so-called A-terms.
The A-terms among left-handed and right-handed sleptons 
and Higgs scalar fields are obtained in the gravity mediation 
as \cite{Kaplunovsky:1993rd}
\begin{equation}
h_{IJ} {l}_J {e}_I H_d =  
h^{(Y)}_{IJ}{l}_J {e}_I H_d  + h^{(K)}_{IJ}{l}_J {e}_I H_d ,
\label{eq:A-term}
\end{equation}
where 
\begin{eqnarray}
h^{(Y)}_{IJ} &=& F^{\Phi_k} \langle \partial_{\Phi_k} \tilde{y}_{IJ}
\rangle ,  
\nonumber \\
h^{(K)}_{IJ}{l}_J {e}_I H_d &=& - 
\langle \tilde{y}_{LJ} \rangle {l}_J {e}_I H_d F^{\Phi_k} K^{L\bar{L}}
\partial_{\Phi_k} K_{\bar{L}I}  \\
& &  -
\langle \tilde{y}_{IM} \rangle {l}_J {e}_I H_d F^{\Phi_k} K^{M\bar{M}} 
\partial_{\Phi_k} K_{\bar{M}J}  \nonumber  \\ 
& & -   \langle \tilde{y}_{IJ} \rangle {l}_J {e}_I H_d F^{\Phi_k} K^{H_d}
\partial_{\Phi_k} K_{H_d}, \nonumber  
\label{eq:A-term-2}
\end{eqnarray}
where $K_{H_d}$ denotes the K\"ahler metric of $H_d$.
In addition, $\tilde{y}_{IJ}$ denotes effective Yukawa couplings 
and in our model it corresponds to 
\begin{eqnarray}
\tilde{y}_{IJ}
 = 
y_1^l 
\begin{pmatrix}\alpha_1  & 0 & 0 \\ 
           0  & \alpha_1  &  0  \\
                 0  & 0 & \alpha_1   \\
 \end{pmatrix} 
+y_2^l  
\begin{pmatrix} \omega\alpha_2-\alpha_3 & 0 & 0 \\ 
               0    & \omega^2\alpha_2-\omega^2\alpha_3 &   0 \\
                 0 & 0 &  \alpha_2-\omega\alpha_3 \\
 \end{pmatrix} .
\label{ME}
 \end{eqnarray}
Then, we obtain 
\begin{eqnarray}
h^{(Y)}_{IJ}
 = 
y_1^l 
\begin{pmatrix}\tilde F^{\alpha_1}  & 0 & 0 \\ 
           0  & \tilde F^{\alpha_1}  &  0  \\
                 0  & 0 & \tilde F^{\alpha_1}   \\
 \end{pmatrix} 
+y_2^l  
\begin{pmatrix} \omega \tilde F^{\alpha_2} - 
\tilde F^{\alpha_3} & 0 & 0 \\ 
               0    & \omega^2 \tilde F^{\alpha_2} -
\omega^2 \tilde F^{\alpha_3} &   0 \\
                 0 & 0 &  \tilde F^{\alpha_2} - 
\omega \tilde F^{\alpha_3} \\
 \end{pmatrix},
 \end{eqnarray}
where $\tilde F^{\alpha_i}=F^{\alpha_i}/\alpha_i$.
Because of $\tilde F^{\alpha_i} = {\cal O}(m_{3/2})$, 
all the diagonal entries of $h^{(Y)}_{IJ}$ may be 
of ${\cal O}(y^l_1 m_{3/2})$.
That would cause a problem.
If $|h_{IJ}/\tilde y_{IJ}|$ is large compared with 
slepton masses, there would be a minimum, where 
charge is broken~\cite{CCBbound}\footnote{
If a decay rate from the realistic minimum to such 
charge breaking minimum is sufficiently small compared 
with the age of the universe, 
that might not be a problem. }.

To avoid this, we require that 
$F^{\alpha_1}/\alpha_1= F^{\alpha_2}/\alpha_2=F^{\alpha_3}/\alpha_3$.
Such a relation can be realized if 
the K\"ahler metric of $\chi_i$ for $i=1,2,3$ is the same and 
the non-perturbative superpotential leading to SUSY breaking 
does not include $\chi_{1,2,3}$.
In this case, we obtain  $F^{\alpha_i}/\alpha_i=m_{3/2}$ 
for $i=1,2,3$.
Then, we obtain 
\begin{equation}
\label{eq:A-Y}
h^{(Y)}_{IJ} = \tilde y_{IJ} m_{3/2} ,
\end{equation}
that is, $h^{(Y)}_{11}={\cal O}(m_{3/2}m_e/m_{\tau})$ 
and $h^{(Y)}_{22} ={\cal O}(m_{3/2}m_\mu /m_{\tau})$.

Next, we estimate $h^{(K)}_{IJ}$.
When we neglect correction terms and use the lowest level 
of K\"ahler potential (\ref{eq:Kahler}), we obtain 
\begin{equation}
h^{(K)}_{IJ} = \tilde y_{IJ} A_0 ,
\end{equation}
where $A_0 ={\cal O}(m_{3/2})$.
Furthermore, we take into account corrections to 
the K\"ahler potential including $\chi_i$, and 
we obtain 
\begin{eqnarray}
h^{(K)}_{IJ} v_d
 =
\begin{pmatrix}m_e A_0 +{\cal O}(m_e \alpha_6^2 m_{3/2})  & 
{\cal O}(m_\mu \alpha^2_6 m_{3/2}) & 
{\cal O}(m_\tau \alpha^2_6 m_{3/2}) \\ 
{\cal O}(m_\mu \alpha^2_6 m_{3/2})  & 
m_\mu A_0 +{\cal O}(m_\mu \alpha^2_6 m_{3/2})    &  
{\cal O}(m_\tau \alpha^2_6 m_{3/2}  \\
 {\cal O}(m_\tau \alpha^2_6 m_{3/2}  & 
{\cal O}(m_\tau \alpha^2_6 m_{3/2} & 
m_\tau A_0  +{\cal O}(m_\tau \alpha^2_6 m_{3/2}^2)   \\
 \end{pmatrix} .  
\end{eqnarray}
This structure does not change except replacing 
$A_0$ by $A_0 + m_{3/2}$ when we include $h^{(Y)}_{IJ}$ (\ref{eq:A-Y}). 
Then, our model predicts  
$h_{12}v_d/m_{3/2}^2 = {\cal O} (\alpha^2_6 m_{\mu}/m_{3/2})$.
This ratio is constrained less than ${\cal O}(10^{-6})$  by 
the $\mu \rightarrow e \gamma$ experiments
when the slepton mass is equal to 100 GeV.
That is, the parameter region with $\alpha^2_6 \leq {\cal O}(10^{-3})$ 
is favorable.
Thus, our parameter region $\alpha_6 \leq 10^{-2}$ in 
the previous section is favorable again from 
the FCNC constraints on the A-terms.

\section{Summary and discussion}
We have presented the  flavor model for the lepton mass matrices 
by using the discrete symmetry $\Delta (54)$, which could be 
originated from heterotic string orbifold models or 
magnetized/intersecting D-brane models. 
The left-handed leptons, the right-handed charged leptons 
and the right-handed neutrinos are assigned to be $3_1^{(1)}$, 
$3_2^{(2)}$ and $1_1+2_1$, respectively. 
We introduce  gauge singlets $\chi_1$, $(\chi_2, \chi_3)$,
$(\chi_4, \chi_5)$, and $(\chi_6,\chi_7,\chi_8)$, which are assigned to be 
$1_2$, $2_1$, $2_1$, and $3_1^{(2)}$ of the $\Delta(54)$ representations, 
respectively.

The discrete symmetry reduces  fine tuning  to
get the tri-bimaximal mixing for arbitrary neutrino masses.
However, some fine tuning is implicitly introduced
in vacuum alignments of scalar fields if those are not 
guaranteed in our model. 
Therefore, we should discuss the origin of vacuum alignments.
One way is to analyze the scalar potential.
Unfortunately,  
the scalar potential is very complicated in our model 
since  there are nine scalar fields which develop their VEVs.
We can only 
say that our desired VEVs are  just one of solutions 
to realize the potential minimum. 
We  can also discuss  new methods\cite{Kobayashi:2008ih,Seidl}
in the extra  dimensional theory,
which  may naturally lead  desired vacuum alignments.
Details will be studied elsewhere.

In our model, we predict the upper bound $0.07$ for $\sin\theta_{13}$.
The  magnitudes of $\sin\theta_{23}$ could  be deviated from the bi-maximal mixing
 considerably, but  $\sin\theta_{12}$  is hardly deviated from
$1/\sqrt3$.

We have also studied SUSY breaking terms.
It is the prediction of our flavor model that
three families of left-handed and right-handed slepton masses 
are almost degenerate.
Our model leads to 
smaller values of FCNCs than the present experimental bounds.

\vspace{1cm}
\noindent
{\bf Acknowledgement}

The work of H.I. is supported by Grand-in-Aid for Scientific Research,
No.21.5817 from the Japan Society of Promotion of Science.
T.~K.\/ is supported in part by the
Grant-in-Aid for Scientific Research, No.~20540266,
and the Grant-in-Aid for the Global COE Program 
"The Next Generation of Physics, Spun from Universality 
and Emergence" from the Ministry of Education, Culture,
Sports, Science and Technology of Japan.
The work of M.T. is  supported by the
Grant-in-Aid for Science Research, No. 21340055,
from the Ministry of Education, Culture,
Sports, Science and Technology of Japan.


\newpage
\noindent
{\bf \LARGE Appendix}
\appendix
\section{Derivation  mixing angles from  mass matrix}
We show analytic expressions for the mixing matrix.
By the unitary transformation,
the neutrino mass matrix $M_\nu$ in Eq.(\ref{Matrix1}) becomes
\begin{eqnarray}
\begin{split}
 \tilde M_\nu=U_\text{tri}^\dagger M_\nu U_\text{tri}
=&
 \begin{pmatrix}m_1+\frac32b\delta   & 0 
& \frac{3\omega^2-3\omega}{2\sqrt3}b\delta \\ 
                   0    & m_2   &0    \\
        \frac{3\omega^2-3\omega}{2\sqrt3}b\delta  & 0  
& m_3-\frac32b\delta    \\
 \end{pmatrix}
,
\end{split}
\end{eqnarray}
where
\begin{eqnarray}
U_\text{tri}=
 \begin{pmatrix}2/\sqrt6   & 1/\sqrt3 & 0 \\ 
                   -1/\sqrt6    & 1/\sqrt3   &-1/\sqrt2    \\
                   -1/\sqrt6  & 1/\sqrt3  & 1/\sqrt2    \\
 \end{pmatrix}.
\end{eqnarray}

Therefore,  the deviation from the tri-bimaximal mixing 
can be expressed by diagonalizing $2\times2$ matrix. 
The matrix $\tilde M_{\nu}$ can be diagonalized by
\begin{eqnarray}
\begin{split}
& \begin{pmatrix}m_1+\frac32b\delta   & 0 
& \frac{3\omega^2-3\omega}{2\sqrt3}b\delta \\ 
                   0    & m_2   &0    \\
                  \frac{3\omega^2-3\omega}{2\sqrt3}b\delta  & 0  
& m_3-\frac32b\delta    \\
 \end{pmatrix}
=P^{-1}V_{13} \begin{pmatrix}m_1'   & 0 & 0 \\ 
                   0    & m_2   &0    \\
                   0  & 0  & m_3'    \\
 \end{pmatrix}
V_{13}^TP^{-1},
\end{split}
\end{eqnarray}
where
\begin{eqnarray}
V_{13}=
 \begin{pmatrix}\cos\theta   & 0 &  - \sin\theta \\ 
                   0    & 1   &0    \\
                 \sin\theta  & 0  & \cos\theta    \\
 \end{pmatrix},
\quad
 P=
 \begin{pmatrix}e^{ip_1}   & 0 & 0 \\ 
                   0    & 1   &0    \\
                   0  & 0  & e^{ip_2}    \\
 \end{pmatrix}.
\end{eqnarray}
We introduce new phase  parameters as follows:
\begin{eqnarray}
m_a= m_1+\frac32b\delta=|m_a|e^{i\mu_a},
\quad
m_b= m_3-\frac32b\delta=|m_b|e^{i\mu_b},
\quad
b=|b|e^{i\beta},
\quad
\delta=|\delta|e^{i\xi}.
\end{eqnarray}
Then, we find the phase difference $p_1-p_2$ and an additional mixing angle
$\theta$ as
\begin{eqnarray}
\begin{split}
&\tan(p_1-p_2)
=\frac{-|m_a|\sin(\mu_a-\pi/2-\beta-\xi)+|m_b|\sin(\mu_b-\pi/2-\beta-\xi)}
{|m_a|\cos(\mu_a-\pi/2-\beta-\xi)+|m_b|\cos(\mu_b-\pi/2-\beta-\xi)},
\\
&\tan(2\theta)
=\frac{-3|b||\delta|}
{|m_a|\cos(\mu_a-\pi/2-\beta-\xi+p_1-p_2)
-|m_b|\cos(\mu_b-\pi/2-\beta-\xi-p_1+p_2)}, \nonumber
\end{split}
\end{eqnarray}
and neutrino masses as
\begin{eqnarray}
\begin{split}
m_1'&=
c^2|m_a|e^{i(\mu_a+2p_1)}
+s^2|m_b|e^{i(\mu_b+2p_2)}
-3cs|b||\delta|e^{i(\pi/2+\beta+\xi+p_1+p_2)},
\\
m_3'&=
s^2|m_a|e^{i(\mu_a+2p_1)}
+c^2|m_b|e^{i(\mu_b+2p_2)}
+3cs|b||\delta|e^{i(\pi/2+\beta+\xi+p_1+p_2)}.
\end{split}
\end{eqnarray}
Mixing matrix of this situation can be expressed by
\begin{eqnarray}
U= U_\text{tri}
P^{-1}V_{13}=
\begin{pmatrix}\frac{\sqrt2}{\sqrt3}e^{-ip_1}\cos\theta   
& \frac{1}{\sqrt3} & -\frac{\sqrt2e^{-ip_1}}{\sqrt3}\sin\theta \\ 
             -\frac{e^{-ip_1}\cos\theta+\sqrt3e^{-ip_2}\sin\theta}{\sqrt6} 
   & \frac{1}{\sqrt3}   
&\frac{e^{-ip_1}\sin\theta-\sqrt3e^{-ip_2}\cos\theta}{\sqrt6}   \\
             -\frac{e^{-ip_1}\cos\theta-\sqrt3e^{-ip_2}\sin\theta}{\sqrt6}  
& \frac{1}{\sqrt3} 
& \frac{e^{-ip_1}\sin\theta+\sqrt3e^{-ip_2}\cos\theta}{\sqrt6}      \\
 \end{pmatrix}.
\end{eqnarray}

In the same way, we diagonalize another neutrino mass matrix $M_{\nu}$
in Eq.(\ref{Matrix2}) as
\begin{eqnarray}
\begin{split}
\tilde M_{\nu}=U_\text{tri}^\dagger M_\nu  U_\text{tri} =
\begin{pmatrix} m_1+\frac{\delta_1+\delta_2}{3}m_1  
& -\frac{(\delta_1+\delta_2)(m_1+m_2)}{3\sqrt2} 
&\frac{(\delta_1-\delta_2)(m_1+m_3)}{2\sqrt3} \\ 
                   -\frac{(\delta_1+\delta_2)(m_1+m_2)}{3\sqrt2}    
&  m_2+\frac{2(\delta_1+\delta_2)}{3}m_2    
&-\frac{(\delta_1-\delta_2)(m_2+m_3)}{\sqrt6}    \\
                  \frac{(\delta_1-\delta_2)(m_1+m_3)}{2\sqrt3}   
& -\frac{(\delta_1-\delta_2)(m_2+m_3)}{\sqrt6}   
&  m_3+({\delta_1+\delta_2})m_3    \\
 \end{pmatrix}.
\end{split}
\end{eqnarray}
For simplicity, 
we assume that $\delta_1$ and $\delta_2$ are real 
and neglect $\delta_1^2$ and $\delta_2^2$, then
new mass eigenvalues are approximately
\begin{eqnarray}
m_1'\simeq
\frac{3+\delta_1+\delta_2}{3}m_1,
\quad
m_2'
\simeq
\frac{3+2(\delta_1+\delta_2)}{3}m_2, 
\quad
m_3'
\simeq
(1+{\delta_1+\delta_2})m_3.
\end{eqnarray} 
Similarly, mixing matrix to diagonalize $\tilde M_\nu$ 
can be also expressed in terms of $\delta_1$ and $\delta_2$ as
\begin{eqnarray}
U=U_\text{tri}
 \begin{pmatrix}\cos\theta_{12}   & \sin\theta_{12} & 0 \\ 
                   -\sin\theta_{12}    & \cos\theta_{12}   &0    \\
                   0  & 0  & 1    \\
 \end{pmatrix}
 \begin{pmatrix}\cos\theta_{13}   & 0 & \sin\theta_{13} \\ 
                   0    & 1   &0    \\
                   -\sin\theta_{13}  & 0  & \cos\theta_{13}    \\
 \end{pmatrix}
  \begin{pmatrix}1   & 0 & 0 \\ 
                   0    & \cos\theta_{23}   &\sin\theta_{23}    \\
                   0  & -\sin\theta_{23}  & \cos\theta_{23}   \\
 \end{pmatrix}
\end{eqnarray}
where
\begin{eqnarray}
\begin{split}
\theta_{12}
&\simeq-\frac{m_1+m_2}{3\sqrt2(m_2-m_1)}
(\delta_1+\delta_2),
\quad
\theta_{13}
\simeq\frac{m_1+m_3}{2\sqrt3(m_3-m_1)}
(\delta_1-\delta_2),
\\
\theta_{23}
&\simeq-\frac{m_2+m_3}{\sqrt6(m_3-m_2)}
(\delta_1-\delta_2).
\end{split}
\end{eqnarray}

\section{Another $\Delta(54)$ flavor model}

Here, for comparison, 
we study soft SUSY breaking terms derived from 
the $\Delta(54)$ flavor model, which was discussed 
in Ref.~\cite{d54our}.
In this model, the flavor symmetry is $\Delta(54)$, but 
there is no additional $Z_2$ flavor symmetry. 
We introduce gauge singlets, $\chi_i$ 
for $i=1,\cdots,6$, and assignments of 
$\Delta(54)$ representations are shown in Table 2.

\begin{table}[h]
\begin{center}
\begin{tabular}{|c|ccc||c||ccc|}
\hline
              &$(L_e,L_\mu,L_\tau)$ & $(e_e^c,e_\mu^c,e_\tau^c)$ & 
$(N_e^c,N_\mu^c,N_\tau^c)$         &$h_{u(d)}$ &$ \chi_1 $ 
&  $(\chi_{2},\chi_3)$&  
$(\chi_{4},\chi_5,\chi_6)$ \\ \hline
$\Delta(54)$      &$3_1^{(1)}$      & $3_2^{(2)}$   & $3_1^{(2)}$          
& $1_1$ & $1_2$  & $2_1$ & $3_1^{(2)}$    \\
\hline
\end{tabular}
\end{center}
\caption{Assignments of $\Delta(54)$ representations}
\end{table}

We assume that $\chi_i$ develop their VEVs and parameterize 
them as $\alpha_i = \chi_i/\Lambda$.
To realize lepton masses and mixing angles, values of 
parameters are required as $\alpha_{1,2,3} = {\cal O}(10^{-2})$ 
and $\alpha_{4,5} = {\cal O}(10^{-4})- {\cal O}(10^{-3})$.
(See for details Ref.~\cite{d54our}.)

Now, let us study soft SUSY breaking scalar masses.
Both the left-handed and right-handed leptons are $\Delta(54)$ triplets 
in this model, too.
At the lowest order, we obtain the same K\"ahler potential for leptons as
(\ref{eq:Kahler}).
Then, at this level,  the prediction for slepton masses is the same as 
(\ref{eq:soft-mass-1}).
That is, three families of left-handed and right-handed slepton masses 
are degenerate.
Next, we consider the corrections including $\chi_i$.
Since $\alpha_{1,2,3}$ are larger than $\alpha_{4,5}$, 
the corrections including the form $\alpha_{1,2,3}$ are 
important.
We examine which corrections including $\chi_{1,2,3}$
are allowed by the $\Delta(54)$ symmetry.
Then, the resultant slepton masses squared 
have the following corrections in the SCKM basis, 
\begin{eqnarray}
(m_{\tilde L}^2)_{ij} = m_{L}^2 
\left(
  \begin{array}{ccc}
1 + {\cal O}(\alpha_1)  &  {\cal O}(\alpha_1^2) &  {\cal O}(\alpha_1^2) \\ 
{\cal O}(\alpha_1^2)  & 1+ {\cal O}(\alpha_1)   & {\cal O}(\alpha_1^2)  \\ 
{\cal O}(\alpha_1^2)   &  {\cal O}(\alpha_1^2)  & 1 + {\cal O}(\alpha_1)   
\\ 
\end{array} \right),
\label{eq:soft-mass-3-L}
\end{eqnarray}
for the left-handed sleptons, and 
\begin{eqnarray}
(m_{\tilde R}^2)_{ij} = m_{R}^2 
\left(
  \begin{array}{ccc}
1 + {\cal O}(\alpha_1)  &  {\cal O}(\alpha_1^2) &  {\cal O}(\alpha_1^2) \\ 
{\cal O}(\alpha_1^2)  & 1+ {\cal O}(\alpha_1)   & {\cal O}(\alpha_1^2)  \\ 
{\cal O}(\alpha_1^2)   &  {\cal O}(\alpha_1^2)  & 1 + {\cal O}(\alpha_1)   
\\ 
\end{array} \right),
\label{eq:soft-mass-3-R}
\end{eqnarray}
for the right-handed sleptons.
Thus, we obtain $(\Delta_{LL})_{12} = (\Delta_{RR})_{12} = 
{\cal O} (\alpha^2_1)$.
To realize the lepton masses, we need $\alpha={\cal O}(10^{-2})$.
Such a parameter region is also favorable from the FCNC constraint.

Similarly, we can estimate the A-terms.
When we take into account important corrections, 
the A-term matrix is estimated as 
\begin{eqnarray}
h_{IJ} v_d
 =
\begin{pmatrix}m_e A_0 +{\cal O}(m_e \tilde \alpha m_{3/2})  & 
{\cal O}(m_\mu \tilde \alpha^2 m_{3/2}) & 
{\cal O}(m_\tau \tilde \alpha^2 m_{3/2}) \\ 
{\cal O}(m_\mu \tilde \alpha^2 m_{3/2})  & 
m_\mu A_0 +{\cal O}(m_\mu \tilde \alpha m_{3/2})    &  
{\cal O}(m_\tau \tilde \alpha^2 m_{3/2}  \\
 {\cal O}(m_\tau \tilde \alpha^2 m_{3/2}  & 
{\cal O}(m_\tau \tilde \alpha^2 m_{3/2} & 
m_\tau A_0  +{\cal O}(m_\tau \tilde \alpha m_{3/2})   \\
 \end{pmatrix},  
\end{eqnarray}
where $A_0 = {\cal O}(m_{3/2})$ and we have also assumed that 
$F^{\alpha_1}/\alpha_1=F^{\alpha_2}/\alpha_2=F^{\alpha_3}/\alpha_3$ 
as in section 5.
When $m_{3/2} = 100$ GeV, we obtain 
$h_{12}v_d/m_{3/2}^2 = {\cal O} (10^{-7})$ for 
 $\alpha={\cal O}(10^{-2})$.
Thus, the parameter region for $\alpha_i$ is favorable 
again from the FCNC constraint of A-terms.

\end{document}